\documentclass[9pt,twocolumn,twoside]{pnas-new}
\templatetype{pnasresearcharticle} % Choose template 
\title{Mixed-order phase transition in a colloidal crystal}

\author[a,b,1]{Ricard Alert}
\author[a,b,c]{Pietro Tierno} 
\author[a,b]{Jaume Casademunt}

\affil[a]{Departament de F\'{i}sica de la Mat\`{e}ria Condensada, Universitat de Barcelona, Av. Diagonal 647, 08028 Barcelona, Spain}
\affil[b]{Universitat de Barcelona Institute of Complex Systems (UBICS), Universitat de Barcelona, Barcelona, Spain}
\affil[c]{Institut de Nanoci\`{e}ncia i Nanotecnologia, Universitat de Barcelona, Barcelona, Spain}

\leadauthor{Alert} 

\significancestatement{Phase transitions are usually classified in two categories: first-order transitions are discontinuous, and second-order transitions are continuous and exhibit critical behavior. However, the so-called mixed-order transitions combine features of both types, such as being discontinuous yet featuring a diverging correlation length. Such transitions were first foreseen by David J. Thouless in 1969 in spin chains with long-range interactions. Since then, they have been predicted for a number of equilibrium and nonequilibrium systems. Here, we predict and experimentally observe a mixed-order phase transition in a colloidal crystal. Our findings constitute the first experimental realization of an equilibrium mixed-order transition, thus enabling the experimental investigation of the surprising properties of such phase transitions.}

\authorcontributions{Author contributions: R.A. conceived the research and developed the theory. P.T. performed the experiments. J.C. supervised the work. All authors discussed and interpreted results. R.A. wrote the paper, with input from all authors.}
\authordeclaration{The authors declare no conflict of interest.}
\correspondingauthor{\textsuperscript{1}To whom correspondence should be addressed. E-mail: ricard.alert@fmc.ub.edu}


\begin{abstract}
Mixed-or\-der phase transitions display a discontinuity in the order parameter like first-or\-der transitions yet feature critical behavior like se\-cond-or\-der transitions. Such transitions have been predicted for a broad range of equilibrium and nonequilibrium systems, but their experimental observation has remained elusive. Here, we analytically predict and experimentally realize a mixed-order equilibrium phase transition. Specifically, a discontinuous solid-solid transition in a two-di\-men\-sio\-nal crystal of paramagnetic colloidal particles is induced by a magnetic field $H$. At the transition field $H_{\text{s}}$, the energy landscape of the system becomes completely flat, which causes diverging fluctuations and correlation length $\xi\propto | H^2 - H_{\text{s}}^2|^{-1/2}$. Mean-field critical exponents are predicted, since the upper critical dimension of the transition is $d_{\text{u}}=2$. Our colloidal system provides an experimental test bed to probe the unconventional properties of mixed-or\-der phase transitions.
\end{abstract}

\dates{This manuscript was compiled on \today}
\doi{\url{www.pnas.org/cgi/doi/10.1073/pnas.XXXXXXXXXX}}

\begin{document}

% Optional adjustment to line up main text (after abstract) of first page with line numbers, when using both lineno and twocolumn options.
% You should only change this length when you've finalised the article contents.
\verticaladjustment{-2pt}

\maketitle
\thispagestyle{firststyle}
\ifthenelse{\boolean{shortarticle}}{\ifthenelse{\boolean{singlecolumn}}{\abscontentformatted}{\abscontent}}{}

% If your first paragraph (i.e. with the \dropcap) contains a list environment (quote, quotation, theorem, definition, enumerate, itemize...), the line after the list may have some extra indentation. If this is the case, add \parshape=0 to the end of the list environment.

\dropcap{P}aul Ehrenfest put forward the first generic classification of phase transitions in 1933 \cite{Jaeger1998}. Nowadays, phase transitions are usually classified into first-order transitions, which feature a discontinuity in the order parameter and finite fluctuations, and second-order transitions, in which the order parameter changes continuously and the correlation length diverges. However, some transitions do not accomodate to this dichotomous classification. In fact, a number of models have been shown to exhibit phase transitions that combine features of first- and second-order transitions. Particularly, transitions that are discontinuous but have a diverging correlation length have been termed ``mixed-order'' or ``hybrid'' phase transitions. Such transitions have been predicted for equilibrium systems with long-range interactions, for example in models of spin chains \cite{Thouless1969,Dyson1971,Aizenman1988,Bar2014,Bar2014a,AnglesdAuriac2016,Fronczak2016}, DNA denaturation \cite{Bar2014,Bar2014a}, and wetting \cite{Indekeu1993,Robledo1994,Blossey1995}, as well as for nonequilibrium phenomena such as jamming \cite{Henkes2005,Toninelli2006,Schwarz2006}, percolation \cite{Schwarz2006,Sheinman2015,Lee2017d}, cooperative dynamics in networks \cite{Bassler2015,Cai2015,Juhasz2017,Lee2017d}, and turbulence \cite{Sahoo2017}. In most cases, these transitions exhibit nonstandard critical behavior, featuring essential singularities or algebraic divergencies with parameter-dependent critical exponents. Whether mixed-order transitions can take place in equilibrium systems with short-range interactions, and whether they can exhibit standard critical behavior was unclear. Moreover, the experimental observation of mixed-order phase transitions has remained elusive.

Here, we unveil a new mixed-order phase transition in a two-dimensional colloidal crystal. We analytically predict the discontinuous character of the transition and the divergence of the correlation length from a model with nearest-neighbor interactions. In addition, we predict that our transition features mean-field critical exponents. Thus, we show that equilibrium systems with short-range interactions can undergo mixed-order transitions, and that they can exhibit mean-field critical behavior in low dimensionality. Moreover, we experimentally observe the transition, thus providing the first experimental realization of an equilibrium mixed-order phase transition.

\section*{Results}

\begin{figure*}[t]
\begin{center}
\includegraphics[width=\textwidth]{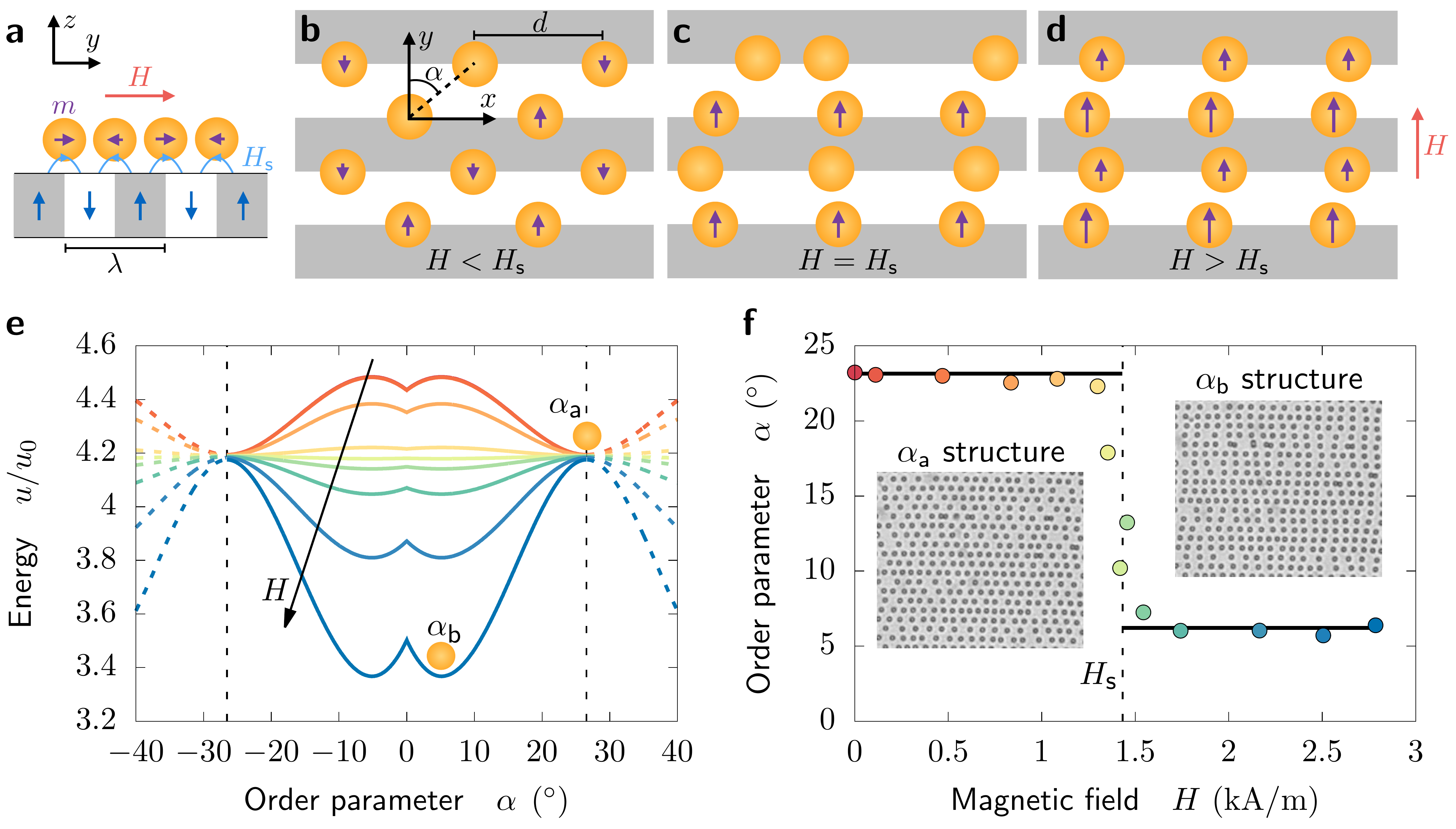}
\caption{\label{fig1}The lands\-cape-in\-ver\-sion phase transition is discontinuous. (\textbf{a}) Sketch of the system (side view). Paramagnetic particles assemble in lines above the walls between oppositely magnetized domains (white and grey) of the substrate. The particles acquire dipolar moments $m\propto H\pm H_{\text{s}}$ due to the superposition of the external and the substrate magnetic fields. (\textbf{b-d}) Sketch of the system (top view), at values of the external magnetic field $H$ lower, equal, or higher than the substrate field $H_{\text{s}}$, respectively. The crystalline order is described by the lattice angle $\alpha$. (\textbf{e}) The energy landscape of the system, Eq. \ref{eq energy}, completely inverts at $H=H_{\text{s}}$, thus inducing a transition between the $\alpha_{\text{a}}$ and $\alpha_{\text{b}}$ crystalline structures. Dashed lines illustrate the periodicity of the potential, with the identification $\alpha_{\text{a}}\leftrightarrow -\alpha_{\text{a}}$, since these angles correspond to the same structure (see \textbf{b}). The energy scale is $u_0\equiv \mu_0\chi^2 a^3 H_\text{s}^2$, and $f\left(\alpha,d/\lambda\right)=\cos^3\alpha (3\cos^2\alpha - 1)+\cos^3\theta (\cos^2\theta - 1)$, with $\theta=\arctan (2d/\lambda-\sqrt{\sec^2\alpha - 1})$. A constant $H^2/H_s^2$ is added to shift the curves for better visualization. (\textbf{f}) Discontinuous transition between the $\alpha_{\text{a}}$ and $\alpha_{\text{b}}$ structures upon increasing the external magnetic field $H$. Points are experimental data and lines indicate the theoretical expectation. The transition field $H_{\text{s}}=1.43\pm 0.04$ kA/m is identified from the midpoint of the order parameter, $\alpha\left(H_{\text{s}}\right)=\left(\alpha_{\text{a}}+\alpha_{\text{b}}\right)/2$.}
\end{center}
\end{figure*}

The colloidal crystal is assembled from a suspension of paramagnetic colloidal particles of radius $a=0.5$ $\mu$m and volume magnetic susceptibility $\chi\approx 1$. The suspension is deposited on a substrate featuring a periodic pattern of magnetic domains in the form of parallel stripes of width $\lambda/2=1.3$ $\mu$m, with consecutive domains having opposite magnetization. The particles then arrange along parallel lines above the domain walls, where the substrate generates a magnetic field $\pm H_{\text{s}}\hat{y}$ that magnetizes particles in consecutive lines in opposite directions (Fig. \ref{fig1}a). Then, dipolar interactions between the particles yield a crystalline ordering characterized by the lattice angle $\alpha$ defined in Fig. \ref{fig1}b.

Upon the application of a uniform in-plane magnetic field $H\hat{y}$, particles on consecutive lines acquire magnetic dipoles $m_i\propto H+H_{\text{s}}$ and $m_j\propto H-H_{\text{s}}$ (Fig. \ref{fig1}a). Thus, their interaction yields a contribution $U_{ij}\propto m_i m_j\propto H_s^2-H^2$ to the energy. In turn, the interactions between particles in the same line yield a constant contribution to the energy, independent of the lattice order parameter $\alpha$. Then, cutting off the dipolar interactions to nearest neighbors, the total interaction energy per particle in a uniform crystal reads \cite{Alert2014}
\begin{multline} \label{eq energy}
u=u_{\text{lines}}+u_\alpha=\\
\bar{u}\left(1+\frac{H^2}{H_{\text{s}}^2}\right)\left(\frac{\lambda/2}{d}\right)^3+\bar{u}\left(1-\frac{H^2}{H_{\text{s}}^2}\right) f\left(\alpha,d/\lambda\right).
\end{multline}
Here $d$ is the mean distance between neighbor particles in the same line (Fig. \ref{fig1}b), and $\bar{u}= 32\pi\mu a^6\chi^2 H_s^2/(9\lambda^3)$, with $\mu\approx 4\pi\times 10^{-7}$ H/m the magnetic permeability of the medium. The shape of $f\left(\alpha,d/\lambda\right)$ can be seen in the energy lanscape (Fig. \ref{fig1}e), which is periodic in $\alpha$, reflecting the crystal periodicity (see Fig. \ref{fig1}b).

For $H<H_s$, the external field $H$ weakens half of the magnetic dipoles and strengthens the other half (Fig. \ref{fig1}b), which lowers the angle-dependent energy contribution $u_{\alpha}$. At $H=H_{\text{s}}$, half of the dipoles are exactly cancelled (Fig. \ref{fig1}c), and $u_\alpha$ vanishes. For $H>H_{\text{s}}$, all the dipoles point in the same direction (Fig. \ref{fig1}d), and $u_\alpha$ changes sign. Therefore, the application of an external field $H>H_s$ causes the energy lanscape to globally invert (Fig. \ref{fig1}e). This induces the so-called landscape-inversion phase transition (LIPT) \cite{Alert2014,Alert2016a}, whereby the crystal experiences a structural transition from the $\alpha_{\text{a}}$ to the $\alpha_{\text{b}}$ structure concomitant with a transition from a ferri- to a ferromagnetic ordering of the induced dipoles (Fig. \ref{fig1}b-f).

The energy landscape in Eq. \ref{eq energy} and Fig. \ref{fig1}e predicts that the equilibrium state is $\alpha_{\text{a}}$ for any field $H<H_{\text{s}}$ and $\alpha_{\text{b}}$ for any field $H>H_{\text{s}}$. Therefore, the order parameter experiences a discontinuity at $H=H_{\text{s}}$, as shown in Fig. \ref{fig1}f. The data confirm this prediction within the experimental accuracy.

Next, we allow for a spatially-dependent, coarse-grained order parameter field $\alpha\left(\vec{r}\right)$ by means of a Ginzburg-Landau effective Hamiltonian \cite{Goldenfeld1992}
\begin{equation}
\mathcal{H}=\rho \int_S \left[u\left(\alpha\right)+\frac{\kappa}{2}\left(\vec{\nabla}\alpha\right)^2\right] d^2\vec{r}.
\end{equation}
Here, $\rho=2/\left(\lambda d\right)$ is the surface density of particles, and $\kappa$ is the spatial coupling coefficient that provides the scale for the energy cost of order-parameter gradients \cite{Cahn1958}. Then, we expand the effective Hamiltonian up to second order in the fluctuations $\delta\alpha\left(\vec{r}\right)=\alpha\left(\vec{r}\right)-\alpha^*$ around the equilibrium value $\alpha^*$ of the order parameter. In terms of the Fourier components of the fluctuation field, it reads
\begin{equation} \label{eq Fourier-Hamiltonian}
\mathcal{H}\approx \frac{N}{2}\sum_{\vec{q}}\left[\bar{u}\left(1-\frac{H^2}{H_{\text{s}}^2}\right)\left.\frac{\partial^2 f}{\partial\alpha^2}\right|_{\alpha^*}+\kappa q^2\right]\left|\delta\tilde{\alpha}_{\vec{q}}\right|^2.
\end{equation}
Here, $N=\rho S$ is the total number of particles, and irrelevant constant terms have been omitted. The structure factor computed from Eq. \ref{eq Fourier-Hamiltonian}, namely within the Gaussian approximation for the fluctuations, takes the Ornstein-Zernike form
\begin{equation} \label{eq OZ}
S\left(q\right)=\left\langle\left|\delta\tilde{\alpha}_{\vec{q}}\right|^2\right\rangle=\frac{R^{-2}}{q^2+\xi^{-2}},
\end{equation}
with a Debye screening length $R=\sqrt{N\kappa/(2k_BT)}$ and a correlation length
\begin{equation} \label{eq correlation-length}
\xi=\left[\frac{\kappa}{\bar{u} \left(1-H^2/H_{\text{s}}^2\right) \left.\partial_\alpha^2 f\right|_{\alpha^*}}\right]^{1/2}.
\end{equation}
This result shows that the correlation length diverges at the transition point $H=H_{\text{s}}$ as $\xi\propto |H^2-H_{\text{s}}^2|^{-1/2}$. The structure factor also diverges at long wavelengths with the mean-field exponent $\eta=0$, defined as $S\left(q,H_{\text{s}}\right)\propto q^{\eta-2}$. Therefore, despite being discontinuous, the LIPT features critical behavior, and hence it has a mixed-order character. Weakly first-order phase transitions, characterized by a small jump of the order parameter, can show apparent signs of criticality. As ordinary first-order transitions, these transitions result from odd terms in the Landau expansion of the free energy, and hence they exhibit hysteresis. In contrast, the LIPT does not exhibit hysteresis (Fig. \ref{fig1}f), and it features genuine critical behavior.

The critical behavior of the LIPT is due to the prefactor $1-H^2/H_{\text{s}}^2$ in the $u_\alpha$ term of the energy, Eq. \ref{eq energy}. This prefactor causes the energy landscape to become completely flat at the transition point (Fig. \ref{fig1}e), which results in the diverging fluctuations and correlation length. In contrast, the Landau free energy for second-order transitions becomes only locally flat at the critical point. In Ginzburg-Landau theory, the Ginzburg criterion identifies the so-called upper critical dimension $d_{\text{u}}$ below which fluctuations are relevant and mean-field scaling fails \cite{Goldenfeld1992}. Consequently, critical exponents depart from their mean-field values in systems of dimensionality $d<d_{\text{u}}$. Whereas $d_u=4$ for Ising-like transitions, application of the Ginzburg criterion to our energy landscape shows that the upper critical dimension of the LIPT is $d_{\text{u}}=2$ (see SI Text). Hence, mean-field critical exponents are expected in our two-dimensional system.

\begin{figure}
\begin{center}
\includegraphics[width=\columnwidth]{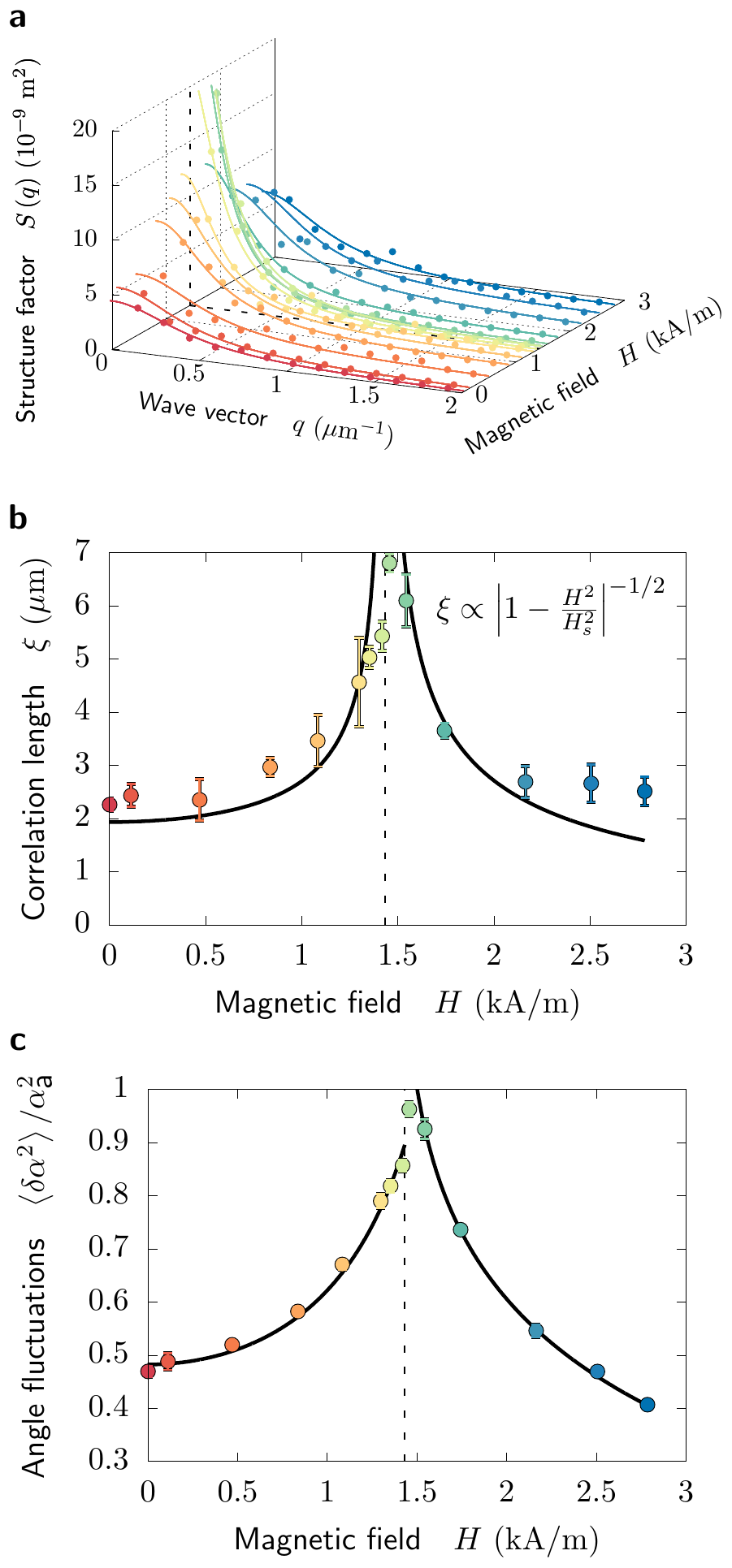}
\caption{\label{fig2}The lands\-cape-in\-ver\-sion phase transition features critical behavior. (\textbf{a}) The structure factor presents a long-wavelength divergence close to the transition point. Points are experimental data and lines are fits of the Ornstein-Zernike form in Eq. \ref{eq OZ}. (\textbf{b}) The correlation length of the order-parameter fluctuations, extracted from the fits of Eq. \ref{eq OZ} to the structure factor, diverges at the transition point. Lines show the predicted scaling, with a fit of the prefactor. (\textbf{c}) The amplitude of the order-parameter fluctuations also diverges at the transition point. Note that angle fluctuations become of the order of the maximum lattice angle $\alpha_{\text{a}}$ (see Fig. \ref{fig1}) close to the critical point. Lines are fits to the prediction, Eq. \ref{eq fluctuations-amplitude} with Eq. \ref{eq correlation-length}, with three fitting parameters: $\kappa,b,N$.}
\end{center}
\end{figure}

To test our predictions, we experimentally measured the structure factor at different values of the external field $H$. Fig. \ref{fig2}a shows the expected divergence of the structure factor at long wavelengths close to the transition point. By fitting Eq. \ref{eq OZ} to the experimental data, we extract the correlation length at different magnetic fields, which is shown to diverge at the transition point in Fig. \ref{fig2}b. The data feature a tendency similar to the predicted mean-field scaling, Eq. \ref{eq correlation-length}. However, the critical exponent cannot be reliably measured in our setup, since thermal fluctuations and the presence of defects limit the size of the crystals. 

We also measured the amplitude of the order-parameter fluctuations, which also diverges at the transition point, as shown in Fig. \ref{fig2}c. In this case, the theoretical prediction is
\begin{multline} \label{eq fluctuations-amplitude}
\left\langle\delta\alpha^2\right\rangle=\sum_{\vec{q}}\left\langle\left|\delta\tilde{\alpha}_{\vec{q}}\right|^2\right\rangle\approx \frac{R^{-2}}{\left(2\pi\right)^2}\int_0^{2\pi} d\theta \int_{\pi/L}^{\pi/b}\frac{q\,dq}{q^2+\xi^{-2}}\\
=\frac{R^{-2}}{4\pi}\ln\left(\frac{1+\left(\pi\xi/b\right)^2}{1+\left(\pi\xi/L\right)^2}\right),
\end{multline}
where $b\sim d$ is a microscopic cutoff, and $L\sim\sqrt{S}=\sqrt{N\rho}$ is the system size. Then, the divergence of the correlation length entails the logarithmic divergence of the fluctuations in Eq. \ref{eq fluctuations-amplitude}, as indicated in Fig. \ref{fig2}c. Therefore, like in a conventional critical point, both the fluctuations and the correlation length diverge.

\section*{Discussion}

To conclude, we have shown that a two-dimensional colloidal crystal features a mixed-order transition, as we predicted from a model with nearest-neighbor interactions. Despite experiencing a discontinuous jump, the order parameter displays unbounded fluctuations and a diverging correlation length at the transition point. The divergence stems from the flattening of the energy landscape at the transition point, which is due to the landscape-inversion mechanism of the transition (LIPT) \cite{Alert2014,Alert2016a}. In addition, our transition features an upper critical dimension equal to the system dimensionality, implying mean-field critical exponents. This result calls for further theoretical and experimental investigations, including assessing the universality of the critical behavior in other systems exhibiting the LIPT scenario \cite{Carstensen2015}.

In all, our findings prove that mixed-order phase transitions can take place in an equilibrium system with short-range interactions, exhibiting mean-field critical exponents in two dimensions. Hence, we establish a paradigmatic example of such transitions. Moreover, we provide a colloidal model system \cite{Li2016} to experimentally explore the properties of mixed-order phase transitions, in particular with regard to their dynamics \cite{Alert2016a}.

\matmethods{
The magnetic substrate is a ferrite garnet film (FGF) grown by dipping liquid phase epitaxy on a $0.5$ mm-thick gadolinium-gallium garnet substrate \cite{Tierno2009}. The FGF was coated with a $1.5$ $\mu$m-thick layer of a positive Photoresist AZ-1512 (Microchem, Newton, MA) by spin coating (Spinner Ws-650Sz, Laurell) and UV irradiation (Mask Aligner MJB4, SUSS Microtec).

A monodisperse (coefficient of variation $\sim 3\%$) suspension of spherical paramagnetic colloidal particles (Dynabeads MyOne, Dynal) was diluted in highly deionized water (Milli-Q, Millipore), and then deposited on top of the substrate. The particles sediment by density mismatch and form a monolayer above the substrate due to the balance between the magnetic attraction with the Bloch walls of the FGF and electrostatic repulsion with the polymer coating. After $15$ min of sedimentation and equilibration, the system was subjected to a static in-plane magnetic field generated by custom-made Helmholtz coils with the axis parallel to the substrate plane, and perpendicular to the Bloch walls. The coils are connected to a DC power supply (TTi-EL302Tv), and they generate a constant and uniform magnetic field over the sample region ($\sim 0.5$ cm$^2$) for the duration of the experiment, as we checked with a Teslameter. After $5$ min of further equilibration, $15$ min-long videos are recorded at $75$ Hz over an area of $140\times 105$ $\mu$m$^2$ by an upright optical microscope (Eclipse Ni, Nikon) with a 100X, 1.3NA objective, and a 0.45TV lens connected to a CCD camera (Balser Scout scA640-74fc). The good agreement between the predicted and the experimental crystalline structures shows that thermal equilibrium is reached in the samples. Further verification is detailed in previous works \cite{Alert2014,Alert2016a}.

Upon filtering the videos to optimize brightness and contrast, particle positions are tracked by a custom-made Matlab program based on the Crocker \& Grier image processing code \cite{Crocker1996}. From particle positions, a lattice angle $\alpha_i$ is assigned to each particle according to its definition in Fig. \ref{fig1}b, so that an order parameter field $\alpha\left(\vec{r},t\right)$ is obtained. From it, the equilibrium order parameter is computed as $\alpha=\langle\langle\alpha\left(\vec{r},t\right)\rangle_{\vec{r}}\rangle_t$, where the external average runs over $30$ snapshots of the system obtained every $30$s. Then, the correlation function
\begin{equation}
C\left(\vec{r}\right)=\frac{\left\langle\left\langle\delta\alpha\left(\vec{r}\,'\right)\delta\alpha\left(\vec{r}\,'+\vec{r}\right)\right\rangle_{\vec{r}\,'}\right\rangle_t-\left\langle\left\langle \delta\alpha\left(\vec{r}\,'\right)\right\rangle_{\vec{r}\,'}\right\rangle_t^2}{\left\langle\left\langle\delta\alpha^2\left(\vec{r}\,'\right)\right\rangle_{\vec{r}\,'}\right\rangle_t-\left\langle\left\langle \delta\alpha\left(\vec{r}\,'\right)\right\rangle_{\vec{r}\,'}\right\rangle_t^2}
\end{equation}
is computed from the order-parameter fluctuations field $\delta\alpha\left(\vec{r}\right)=\alpha\left(\vec{r}\right)-\left\langle\alpha\left(\vec{r}\,'\right)\right\rangle_{\vec{r}\,'}$. Finally, the structure factor is obtained as
\begin{equation}
S\left(q\right)=\int_S C\left(\vec{r}\right)e^{-i\vec{q}\cdot\vec{r}}\,d^2\vec{r}.
\end{equation}
}

\showmatmethods % Display the Materials and Methods section

\acknow{We thank Tom H. Johansen for the ferrite garnet film. R.A. acknowledges support from Fundaci\'{o} ``La Caixa''. P.T. acknowledges the European Research Council under project No. 335040. All authors acknowledge the MINECO under project FIS2016-78507-C2-2-P and Generalitat de Catalunya under project 2014-SGR-878.}

\showacknow % Display the acknowledgments section

% \pnasbreak splits and balances the columns before the references.
% If you see unexpected formatting errors, try commenting out this line
% as it can run into problems with floats and footnotes on the final page.
\pnasbreak

% Bibliography
\bibliography{Criticality}

\end{document}